\PassOptionsToPackage{sort&compress}{NJDnatbib}
\documentclass[AMA,STIX2COL]{MRM}
\articletype{Research Article}
\received{xx July 2026}
\revised{xx xxx xxxx}
\accepted{xx xxx xxxx}
\usepackage{amsmath,amssymb,amsfonts}
\usepackage{mathtools}
\usepackage{array}
\usepackage{booktabs}
\usepackage{etoolbox}

\newcommand{\method}{MOSAIC}
\newcommand{\methodAblationSen}{w/o maps}
\newcommand{\methodAblationSenEcho}{w/o maps$+$encoding}
\newcommand{\echo}{echo}

\renewcommand{\eqref}[1]{(\ref{eq:#1})}

\newcommand{\tabref}[1]{Table~\ref{tab:#1}\!\!\!}
\newcommand{\tabsref}[1]{Tables~\ref{tab:#1}\!\!\!}
\newcommand{\tabnoref}[1]{\ref{tab:#1}\!\!\!}

\newcommand{\figref}[1]{Figure~\ref{fig:#1}\!\!\!}
\newcommand{\figsref}[1]{Figures~\ref{fig:#1}\!\!\!}
\newcommand{\fignoref}[1]{\ref{fig:#1}\!\!\!}

\renewcommand{\Tilde}{\widetilde}
\renewcommand{\Hat}{\widehat}

\renewcommand{\vec}[1]{\ensuremath{\boldsymbol{#1}}}
\newcommand{\Tvec}[1]{\ensuremath{\Tilde{\boldsymbol{#1}}}}
\newcommand{\Hvec}[1]{\ensuremath{\Hat{\boldsymbol{#1}}}}

\newcommand{\Real}{{\mathbb{R}}}
\newcommand{\Complex}{{\mathbb{C}}}

\DeclareMathOperator*{\argmin}{arg\,min}

\begin{document}
\title{\method: A Self-supervised Dynamic Multi-encoding Reconstruction Framework for 3D Late Gadolinium Enhancement MRI}

\author[1]{Muhammad A. Sultan}{\orcid{0009-0007-0185-7400}}
\author[2]{Yingmin Liu}{}
\author[3]{Katherine Binzel}{}
\author[4]{Katarzyna E. Gil}{}
\author[4]{Karolina M. Zareba}{}
\author[1,5]{Rizwan Ahmad}{\orcid{0000-0002-5917-3788}}

\authormark{Sultan \textsc{et al.}}

\address[1]{\orgdiv{Department of Biomedical Engineering},
\orgname{The Ohio State University},
\orgaddress{\city{Columbus}, \state{Ohio}, \postcode{43210}, \country{USA}}}

\address[2]{\orgdiv{Davis Heart and Lung Research Institute},
\orgname{The Ohio State University Wexner Medical Center},
\orgaddress{\city{Columbus}, \state{Ohio}, \postcode{43210}, \country{USA}}}

\address[3]{\orgdiv{Department of Radiology},
\orgname{The Ohio State University Wexner Medical Center},
\orgaddress{\city{Columbus}, \state{Ohio}, \postcode{43210}, \country{USA}}}

\address[4]{\orgdiv{Department of Internal Medicine, Division of Cardiovascular Medicine},
\orgname{The Ohio State University Wexner Medical Center},
\orgaddress{\city{Columbus}, \state{Ohio}, \postcode{43210}, \country{USA}}}

\address[5]{\orgdiv{Department of Electrical and Computer Engineering},
\orgname{The Ohio State University},
\orgaddress{\city{Columbus}, \state{Ohio}, \postcode{43210}, \country{USA}}}

\corres{Rizwan Ahmad, \email{ahmad.46@osu.edu}}

\finfo{\fundingAgency{National Institutes of Health (NIH)} grants
\fundingNumber{R01HL151697} and \fundingNumber{R01EB029957}.}

\abstract[Abstract]{
\section{Purpose} To develop and evaluate a self-supervised dynamic reconstruction framework for highly undersampled dual-\echo\ three-dimensional late gadolinium enhancement (3D LGE) MRI.
\section{Methods} \method\ jointly models multi-\echo\ image content, coil sensitivity maps, and beat-specific nonrigid motion directly from acquired undersampled data, without requiring fully sampled training datasets or accurate precomputed sensitivity maps. 
Unlike existing methods that bin the acquired data into different motion states, with or without motion compensation, \method\ reconstructs a motion-resolved 3D LGE image from each heartbeat. 
The method was evaluated using digital phantoms with simulated myocardial scars and in vivo animal and human studies. 
\section{Results} In phantom experiments, \method\ achieved higher peak signal-to-noise ratio and structural similarity index measure than low-rank deep image prior reconstruction and ablation variants of \method.
In animal and human studies, \method\ achieved higher blinded expert image-quality scores than inline image-navigated compressed-sensing and low-rank deep image prior reconstructions.
\section{Conclusion} \method\ demonstrated the feasibility of motion-resolved free-breathing dual-\echo\ 3D LGE MRI at acceleration factors exceeding $1,\!000$, with improved detail preservation and artifact suppression relative to the state-of-the-art comparison methods.
}




\maketitle
\clearpage

\section{Introduction}
\label{sec:introduction}
Late gadolinium enhancement (LGE) magnetic resonance imaging (MRI) is the clinical standard for detecting and quantifying myocardial scar and plays an important role in diagnosis, risk stratification, and treatment planning across a broad range of cardiomyopathies.\cite{kim2000lge,piehler2013FB-LGE,toupin2022lge} In clinical practice, LGE is commonly performed using segmented two-dimensional inversion-recovery acquisitions with electrocardiographic gating over multiple breath-holds.\cite{kellman2002PSIR, abdula2014psir} Although widely used, this approach can be challenging in patients with limited breath-hold capacity or arrhythmias and provides limited through-plane resolution. Free-breathing three-dimensional (3D) LGE offers contiguous near-isotropic coverage that can be reformatted in arbitrary planes, improving visualization of small or apical scars and enabling more comprehensive scar assessment.\cite{peters2009lge3d,kino2009threeD,morita2013comparison,morsbach2016comp,pennig2020lge} However, robust free-breathing 3D LGE remains challenging due to limitations of existing motion-compensation techniques.

Early free-breathing 3D LGE methods used diaphragm navigators (dNAV) to prospectively gate data acquisition within a selected respiratory window.\cite{saranathan2004dNAV, nguyen2008dNAV3d} Because data outside this window are not acquired, scan efficiency depends on the subject's breathing pattern, which can lead to prolonged, unpredictable scan times. To improve efficiency, navigator information can also be used retrospectively during reconstruction, where the acquired data are sorted into a small number of respiratory bins and the inter-bin motion is estimated as a part of the reconstruction process.\cite{prieto2015dNAVretro, munoz2020binning, zeilinger2022binning} More recent image-navigated (iNAV) 3D LGE methods acquire a low-resolution image of the heart before the 3D LGE readout in each heartbeat, enabling beat-specific respiratory-motion estimation and substantially improving scan efficiency while providing a more predictable 5- to 10-minute scan time.\cite{bratis2016iNAV-seq, zeilinger2021iNAV, holtackers2022lge, hopman2024dNAVvsiNAV} Nevertheless, current iNAV-based LGE workflows remain susceptible to residual motion artifacts due to intra-bin motion or incomplete beat-specific motion models that estimate only left-right and head-foot translations from the low-resolution coronal iNAV image while leaving anterior-posterior and nonrigid motion unmodeled.\cite{bratis2016iNAV-seq, zeilinger2021iNAV, munoz2020binning, zeilinger2022binning}

To improve visualization of myocardial scar adjacent to epicardial fat, multi-\echo\ Dixon-based 3D LGE acquisitions have also been explored to provide water-only and fat-only images.\cite{ma2008dixon, zeilinger2021iNAV, munoz2020binning, zeilinger2022binning} In multi-\echo\ 3D acquisition, readouts at multiple distinct \echo\ times are acquired in each heartbeat; these readouts share the same motion state but differ in contrast based on their echo time. However, existing multi-\echo\ LGE reconstruction workflows, including iNAV, use the multi-\echo\ data primarily for post-reconstruction fat-water separation and reconstruct each \echo\ independently, thereby not exploiting inter-\echo\ redundancy. 

An alternative to conventional respiratory-compensated LGE reconstruction, which integrates data across respiratory phases into a single motion-corrected image and may remain susceptible to residual artifacts from incomplete motion modeling, is motion-resolved reconstruction, which generates a 3D LGE image for each heartbeat by modeling respiratory motion in a beat-specific manner.
Self-supervised deep learning reconstruction methods have recently shown promise for highly accelerated dynamic MRI, where high-quality, fully sampled reference data for supervised training are generally not available.
Low-rank deep image prior (LR-DIP) combines deep image prior (DIP) with subspace modeling to represent dynamic 2D or 3D cine series using learned spatial and temporal bases.\cite{hamilton2023lrdip2d, hamilton2024lrdip3d} Other related methods include manifold-based approaches such as DISCUS\cite{sultan2025discus} for 2D LGE, and motion-aware formulations that incorporate explicit deformation fields.\cite{zou2022moco-storm, vornehm2025mdip} More recently, ML-DIP combined DIP with manifold learning and low-rank representation of image content and motion for highly accelerated 3D real-time cine reconstruction.\cite{chen2025mldip} However, ML-DIP was developed for 3D real-time cine and is not directly applicable to multi-\echo\ free-breathing 3D LGE.

Reconstructing a 3D LGE image from each heartbeat is a highly ill-posed problem that remains largely unexplored in the context of free-breathing 3D LGE.
With only $\sim$15--25 readout lines acquired per echo per heartbeat (frame), the net acceleration rates can exceed $1,\!000$. 
Unlike 3D real-time cine, where a five-minute scan provides thousands of frames, iNAV-based LGE typically provides only a few hundred frames, leading to substantial data scarcity.
In addition, unlike cine, free-breathing LGE is dominated by respiratory motion, with only residual cardiac variation.
Furthermore, most existing 3D reconstruction methods assume that coil sensitivity maps are available a priori,\cite{hamilton2024lrdip3d, chen2025mldip} for example, from ESPIRiT.\cite{uecker2014espirit} In 3D LGE, however, the extent of the fully sampled auto-calibration signal region from the time-averaged k-space is limited, which can make sensitivity map estimation unreliable.\cite{hamilton2025multifrequency_timeDIP} These considerations motivate a dedicated framework for highly undersampled multi-\echo\ 3D LGE.

In this work, we develop \method\ (MOtion-resolved Shared-bAsis and Integrated Calibration), a reconstruction framework for highly undersampled multi-\echo\ 3D LGE that jointly models beat-specific motion, multi-\echo\ image content, and coil sensitivity maps.
Compared to the existing self-supervised methods used for 3D cine, this work offers two specific technical innovations: 1) joint multi-\echo\ reconstruction using a shared image basis to exploit inter-\echo\ redundancy, and 2) joint estimation of coil sensitivity maps during reconstruction \cite{leynes2024nlinv} to reduce reliance on potentially inaccurate fixed maps estimated from limited calibration data. In contrast to conventional iNAV-based motion-compensated reconstruction methods,\cite{zeilinger2021iNAV, zeilinger2022binning} \method\ reconstructs a 3D image at each heartbeat and therefore avoids respiratory binning or data integration. 
We evaluate the proposed reconstruction framework using digital phantoms and in vivo animal and human datasets, with comparison against extended LR-DIP and inline iNAV-based compressed-sensing reconstruction. 
This study extends our preliminary conference abstract in SCMR \cite{sultan2026mledips} and provides the first detailed description and validation of \method.

\section{Methods}
\subsection{Problem formulation for motion-resolved dual-\echo\ 3D LGE}\label{sec:problem}
In multi-\echo\ free-breathing 3D LGE, a small number of readouts are measured in each heartbeat (frame) at multiple \echo\ times. A nonselective inversion recovery pulse is applied prior to acquisition to provide T1-weighted contrast between normal and infarcted myocardium. The process is repeated for hundreds of frames to provide more uniform coverage of the underlying 3D k-space by collecting complementary samples over time. Rather than sorting the data into respiratory bins, our goal is to recover motion-resolved multi-\echo\ frames.

Let $\vec{X} \coloneq \{\vec{x}_{e}^{(t)}\}_{t=1,e=1}^{T,E}$ denote the multi-\echo\ 3D image series. We define $\vec{X}^{(t)} \coloneq \{\vec{x}_{e}^{(t)}\}_{e=1}^{E}$ as the collection of echo images at frame $t$, and $\vec{X}_e \coloneq \{\vec{x}_{e}^{(t)}\}_{t=1}^{T}$ as the collection of frames for echo $e$. Here, $T$ and $E$ denote the total numbers of frames and echoes, respectively, and $\vec{x}_{e}^{(t)}\in\Complex^{N\times 1}$ is the vectorized 3D image for echo $e$ at frame $t$, with $N=n_1\times n_2\times n_3$ voxels. Similarly, let $\vec{y}_{e}^{(t)}\in\Complex^{M\times 1}$ denote the acquired k-space data for echo $e$ at frame $t$ from $N_c$ receive coils, where $M$ is the total number of acquired k-space samples across all receive coils. 
The forward operator is defined as $\vec{A}_{e,\vec{H}}^{(t)}=\vec{P}_e^{(t)}\vec{F}\vec{H}\in\Complex^{M\times N}$, where $\vec{P}_e^{(t)}$ represents the time- and echo-dependent k-space sampling operator, $\vec{F}$ is the multi-coil 3D Fourier transform, and $\vec{H}\in\Complex^{NN_c\times N}$ contains the coil sensitivity maps arranged along the block diagonals of vertically concatenated $N\times N$ blocks. A conventional regularized least-squares formulation, applied independently to each echo, can then be written as
\begin{equation}
    \Hvec{X}_{e} = \argmin_{\vec{X}_{e}} \sum_{t=1}^T \left\| \vec{A}_{e,\vec{H}}^{(t)}\vec{x}_{e}^{(t)} - \vec{y}_{e}^{(t)} \right\|_2^2 + \lambda\mathcal{R}(\vec{X}_{e}),
    \label{eq:ls}
\end{equation}
where $\mathcal{R}(\cdot)$ imposes spatial and/or temporal structure and $\lambda \geq 0$ controls the strength of regularization.

Direct recovery of $\vec{x}_{e}^{(t)}$ from \eqref{ls} is highly ill-posed because each frame must be acquired within a short temporal footprint. In the present setting, each $\vec{y}_{e}^{(t)}$ contains only $\sim$15--25 readout lines to maintain a temporal footprint below 250 ms, corresponding to acceleration factors exceeding $1,\!000$. Existing 3D LGE methods therefore commonly sort the data collected over several minutes into a small number of respiratory bins or apply rigid-body motion correction to align the data across heartbeats.\cite{munoz2020binning, zeilinger2021iNAV, zeilinger2022binning} These approaches often fail in patients with arrhythmias or irregular respiratory patterns. In this work, we instead seek to recover the full motion-resolved multi-\echo\ series, $\vec{X}$, directly.

\begin{figure*}[!t]
\centering
\includegraphics[width=0.95\textwidth]{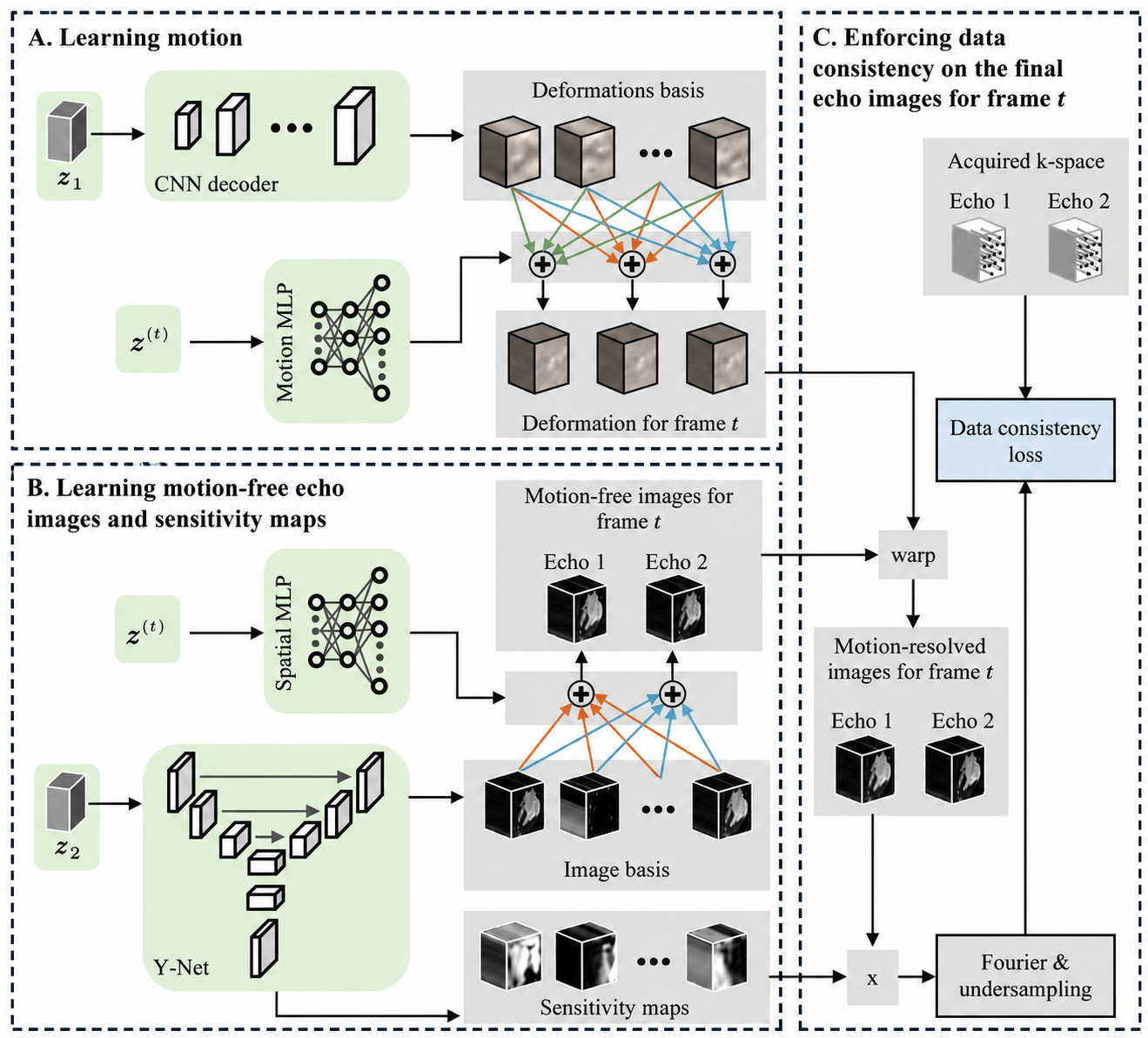}
\caption{Overview of the proposed \method\ framework for dual-\echo\ 3D LGE, showing the flow of information for the $t^{\text{th}}$ frame. (A) Learning motion: A convolutional neural network (CNN) decoder maps a static latent code vector $\vec{z}_1$ to a deformation basis, whose elements are combined using frame-specific coefficients generated by a motion multilayer perceptron (MLP) driven by dynamic latent code vectors $\vec{z}^{(t)}$ to form three frame-specific deformation fields. (B) Learning motion-free echo images and sensitivity maps: A convolutional Y-Net maps a second static latent code vector $\vec{z}_2$ to an image basis and coil sensitivity maps. A spatial MLP generates echo- and frame-specific coefficients to synthesize motion-free dual-\echo\ images from the shared image basis. (C) Enforcing data consistency on the final echo images for frame $t$: The motion-free echo images from (B) are spatially warped with estimated deformation fields from (A) to obtain motion-resolved echo images, which are then multiplied (``x'') with generated sensitivity maps from (B), applied Fourier and sampling operations, and finally made consistent with acquired dual-\echo\ k-space data.}
\label{fig:overview}
\end{figure*}

\subsection{Proposed \method\ reconstruction}
\subsubsection{Framework overview}
To avoid respiratory binning in multi-\echo\ 3D LGE, we propose \method, a motion-resolved shared-basis reconstruction framework with integrated sensitivity-map estimation. Compared with a recently proposed scan-specific self-supervised reconstruction method,\cite{chen2025mldip} \method\ introduces two key features: joint multi-\echo\ reconstruction and joint sensitivity-map estimation.

As shown in \figref{overview}, \method\ consists of three coupled components: (A) beat-specific motion modeling from deformation basis, (B) multi-\echo\ image synthesis from a shared basis and sensitivity map generation, and (C) forward-model-based data consistency. The deformation basis, image basis, and sensitivity maps are shared across the dynamic series, whereas frame- and \echo-specific coefficients used to combine basis elements capture temporal variation. The resulting motion-resolved \echo\ images are constrained to match the acquired multi-\echo\ k-space data, allowing \method\ to exploit inter-\echo\ redundancy while avoiding respiratory binning and reducing reliance on precomputed sensitivity maps.


\subsubsection{Bases and sensitivity map generation}
Two CNN-based generators produce the image basis, deformation basis, and sensitivity maps. A CNN-based decoder, $\mathcal{F}_{\vec{\alpha}_1}$, parameterized by $\vec{\alpha}_1$, maps a static latent code vector $\vec{z}_{1}$ to a deformation basis $\vec{D}$, i.e., $\vec{D} = \mathcal{F}_{\vec{\alpha}_1}(\vec{z}_1)$. Here $\vec{D}= [\vec{d}_1,\vec{d}_2,\dots,\vec{d}_{L_1}]\in \Real^{N\times L_1}$, where $\vec{d}_i\in\Real^{N\times 1}$ denotes the $i^{\text{th}}$ element and $L_1$ is the number of deformation basis elements. A CNN-based Y-Net, $\mathcal{F}_{\vec{\alpha}_2}$, parameterized by $\vec{\alpha}_2$, maps a second static latent code vector $\vec{z}_{2}$ to image basis $\vec{B}$ and sensitivity maps $\vec{H}$, i.e., $\vec{B}, \vec{H} = \mathcal{F}_{\vec{\alpha}_2}(\vec{z}_2)$. Here $\vec{B} = [\vec{b}_1,\vec{b}_2,\dots,\vec{b}_{L_2}]\in\Complex^{N\times L_2}$, where $\vec{b}_i\in\Complex^{N\times 1}$ denotes the $i^{\text{th}}$ element and $L_2$ is the number of basis elements, and block diagonal matrix $\vec{H}\in \Complex^{NN_c\times N}$ carries sensitivity maps along the diagonals of each vertically concatenated $N\times N$ block. Because both generators are driven by static latent vectors, they produce a single deformation basis, a single image basis, and one set of sensitivity maps shared across the full dynamic series.


\subsubsection{Frame-specific coefficient generation}
Temporal variation is modeled using two lightweight fully connected networks driven by frame-specific (beat-specific) latent vectors $\vec{z}^{(t)}\in\Real^{K\times 1}$ as input, where $K$ is the latent dimensionality. We define the collection of dynamic latent code vectors across all $T$ frames as $\vec{Z} \coloneq \{\vec{z}^{(t)}\}_{t=1}^T$. The motion MLP, $\mathcal{F}_{\vec{\alpha}_3}$, parameterized by $\vec{\alpha}_3$, maps $\vec{z}^{(t)}$ to a coefficient matrix $\vec{W}_{d}^{(t)}$, i.e., $\vec{W}_{d}^{(t)} = \mathcal{F}_{\vec{\alpha}_3}(\vec{z}^{(t)})$, where $\vec{W}_{d}^{(t)} \in \Real^{L_1 \times 3}$ combines the deformation basis elements into the frame-specific three-directional deformation fields $\vec{\Phi}^{(t)}=\vec{D}\vec{W}_{d}^{(t)}\in \Real^{N\times 3}$. Likewise, the spatial MLP, $\mathcal{F}_{\vec{\alpha}_4}$, parameterized by $\vec{\alpha}_4$, maps $\vec{z}^{(t)}$ to an echo- and frame-specific coefficient matrix $\vec{W}_{b}^{(t)}$, i.e., $\vec{W}_{b}^{(t)} = \mathcal{F}_{\vec{\alpha}_4}(\vec{z}^{(t)})$, where $\vec{W}_{b}^{(t)} \in \Complex^{L_2 \times E}$ combines the image basis elements into motion-free multi-\echo\ images $\vec{S}^{(t)}=\vec{B}\vec{W}_b^{(t)}\in\Complex^{N\times E}$, with the $e^{\text{th}}$ echo image defined as the $e^{\text{th}}$ column of $\vec{S}^{(t)}$, i.e., $\vec{s}_e^{(t)} \coloneq \vec{S}^{(t)}[:,e]\in\Complex^{N\times 1}$. An estimate of the final echo- and frame-specific image, $\Tvec{x}_e^{(t)}\in\Complex^{N\times 1}$, is generated by warping $\vec{s}_e^{(t)}$ with $\vec{\Phi}^{(t)}$, i.e.,
\begin{equation}
\Tvec{x}_e^{(t)} = \mathcal{W}\left(\vec{s}_e^{(t)}, \vec{\Phi}^{(t)}\right),
\label{eq:warp}
\end{equation}
where $\mathcal{W}\left(\cdot, \cdot\right)$ denotes spatial warping \cite{jaderberg2015warp} of the first argument using the deformation fields provided in the second argument after both $\vec{s}_e^{(t)}$ and $\vec{\Phi}^{(t)}$ have been reshaped into 3D arrays. Note, for a given frame, the same deformation field is jointly applied across all echoes.



\subsubsection{Objective function}
The output of \eqref{warp} is subjected to the forward operator based on the Y-Net-generated sensitivity maps, leading to the following data consistency loss for the $t^{\text{th}}$ frame and $e^{\text{th}}$ echo, 

\begin{equation}
\mathcal{L}_{\mathrm{dc}}(e,t)= \Big\|\vec{A}_{e,\vec{H}}^{(t)} \underbrace{\mathcal{W}\big(
\vec{s}^{(t)}_{e}, \vec{\Phi}^{(t)}\big)}_{\Tvec{x}^{(t)}_{e}} -\vec{y}_e^{(t)} \Big\|_2^2.
\label{eq:dc}
\end{equation}

Since frame-specific 3D LGE recovery is highly ill-posed, we augment the data consistency loss in \eqref{dc} with regularization terms, yielding the following composite loss function.


\begin{equation}
\begin{aligned}
\mathcal{L}_{\mathrm{tot}} = 
&\sum_{t=1}^{T} \sum_{e=1}^{E}
\Big[
\mathcal{L}_{\mathrm{dc}}(e,t) 
+ \lambda_1\mathcal{R}_1(\vec{s}_e^{(t)})
\Big] 
+ \lambda_2 \sum_{t=1}^{T} \mathcal{R}_2(\vec{\Phi}^{(t)}) \\
&\quad
+ \lambda_3\mathcal{R}_3(\vec{H})
+ \lambda_4\mathcal{R}_4(\vec{H}-\vec{H}_0),
\end{aligned}
\label{eq:loss}
\end{equation}
where $\mathcal{R}_1$ is a normalized isotropic total-variation penalty applied to the motion-free echo images along the three spatial directions, $\mathcal{R}_2$ is a normalized first-order finite-difference smoothness penalty applied to the three components of each deformation field, $\mathcal{R}_3$ is a normalized first-order finite-difference smoothness penalty applied to the estimated coil sensitivity maps, and $\mathcal{R}_4$ is a normalized quadratic penalty that constrains the estimated coil sensitivity maps $\vec{H}$ to remain close to the ESPIRiT-based initialization $\vec{H}_0$.


The parameters of the four subnetworks ($\vec{\alpha}_1$, $\vec{\alpha}_2$, $\vec{\alpha}_3$, and $\vec{\alpha}_4$) and the three sets of latent code vectors ($\vec{z}_1$, $\vec{z}_2$, and $\vec{Z}$) are estimated jointly by solving
\begin{equation}
\begin{aligned}
\Hvec{\alpha}_1, \ldots ,\Hvec{\alpha}_4,
\Hvec{z}_1, \Hvec{z}_2, \Hvec{Z} = \argmin_{\vec{\alpha}_1, \ldots, \vec{\alpha}_4, \vec{z}_1, \vec{z}_2, \vec{Z}}\;
\mathcal{L}_{\mathrm{tot}}.
\end{aligned}
\label{eq:opt}
\end{equation}

Once trained, $\Hvec{x}_e^{(t)}$ can be inferred by passing the optimized latent code vectors ($\Hvec{z}_1$, $\Hvec{z}_2$, and $\Hvec{z}^{(t)}$) through the trained networks parameterized by ${\Hvec{\alpha}_1}$, ${\Hvec{\alpha}_2}$, ${\Hvec{\alpha}_3}$, and ${\Hvec{\alpha}_4}$.

\subsection{Implementation and optimization}
\subsubsection{Model configuration}
The proposed framework uses four subnetworks and three sets of latent code vectors. The static latent vector $\vec{z}_1$ was represented as a two-channel real-valued 3D array. The CNN decoder was designed to map $\vec{z}_1$ to spatially smooth deformation basis elements. The static latent vector $\vec{z}_2$ was also represented as a two-channel real-valued 3D array. The Y-Net was designed to map $\vec{z}_2$ to image basis elements and sensitivity maps from a common bottleneck. This Y-Net can be viewed as a single CNN encoder connected to two different CNN decoders.
The image-basis decoder incorporated skip connections from the common encoder, whereas the sensitivity-map decoder used aggressive upsampling layers to promote spatial smoothness in the sensitivity maps.
The frame-specific latent vectors $\vec{z}^{(t)}$ were implemented as $3\times 1$ real-valued vectors and shared across the motion and spatial MLPs. Additional architectural details and code are available at \url{https://github.com/OSU-MR/MOSAIC}. 

\subsubsection{Regularization and initialization}
The numbers of deformation and image basis elements were set to $L_1=9$ and $L_2=8$, respectively. The values of the regularization parameters in \eqref{loss} were selected as $\lambda_1= 0.05$, $\lambda_2= 0.005$, $\lambda_3= 0.01$, and $\lambda_4= 0.01$. To avoid overfitting and to promote smooth deformation fields, a dropout rate of 5\% was used in the CNN decoder. These values were empirically selected using two additional LGE datasets and were kept constant for all cases. 

Unlike the other regularization terms, $\mathcal{R}_4$ was used only to stabilize early training by encouraging the learned sensitivity maps to remain close to the ESPIRiT initialization, $\vec{H}_0$. Because $\vec{H}_0$ may be suboptimal, $\lambda_4$ was set to zero after half of the training iterations. In practice, the initial value of $\lambda_4$ did not have a significant impact on image quality as long as it was large enough to guide early map generation.

\subsubsection{Training and inference}

All subnetworks and latent code vectors were optimized jointly using the Adam optimizer for 20,000 iterations. Each update used a batch of 8 contiguous frames. The learning rate followed a step-decay schedule with a step size of 500 iterations, decreasing from $1\times10^{-3}$ to $5\times10^{-4}$ during training. The total number of learnable parameters, including network parameters and code vectors, was approximately 6 million. After training, the optimized network parameters and code vectors were saved and used to generate a reconstruction of any user-selected frame or set of frames.

\subsection{Experimental studies}
\subsubsection{3D LGE phantom study}
The proposed method was first evaluated using the MRXCAT numerical phantom,\cite{mrxcat_wissmann2014} which enables simulation of realistic 3D cardiac volumes with configurable tissue properties and motion patterns. To assess robustness across subject anatomy and scar presentation, six distinct anatomies (three male, three female) were generated. For each phantom, a myocardial scar was simulated in the left ventricular myocardium, and the scar configurations were varied to span different scar extents and locations.


For each phantom, fat and water images were simulated at an isotropic spatial resolution of 1.25 mm with 36 frames, spanning 36 heartbeats and 8 respiratory cycles. To approximate a 5-minute free-breathing acquisition, each sequence was repeated 8 times, resulting in a total of $T=288$ frames. Fat and water images were converted into a two-\echo\ dataset ($E=2$) using the following simplified Dixon model\cite{ma2008dixon}:
\begin{equation}
\begin{aligned}
\vec{x}_{e}^{(t)}=\vec{x}_W^{(t)} + \vec{x}_F^{(t)} \exp(j\theta_{e}),
\end{aligned}
\label{eq:dixon}
\end{equation}
where $\vec{x}_{W}^{(t)}$ and $\vec{x}_{F}^{(t)}$ denote 3D water and fat images at frame $t$, respectively, $\theta_e$ is the echo-dependent phase, and $\vec{x}_{e}^{(t)}$ is the synthesized echo image. The echo angles were set to $\theta_1=30^\circ$ and $\theta_2=150^\circ$. To reduce computational cost, the imaging volume was cropped to $144 \times 240 \times 86$ voxels along the left-right, anterior-posterior, and head-foot directions, respectively. To better approximate partial-volume blurring observed in acquired images, the simulated images were smoothed using an isotropic 3D Gaussian kernel of size $3 \times 3 \times 3$ voxels with a standard deviation of $0.5$ voxels.

Multi-coil k-space data were generated using 8 receive coils modeled with the Biot-Savart law. Complex-valued white Gaussian noise was added to achieve a signal-to-noise ratio (SNR) of 20 dB. The resulting noisy data were retrospectively undersampled using pseudo-radial sampling (PR4D).\cite{joshi2022sampling} The left-right direction was assigned as the readout dimension ($k_x$), the anterior-posterior direction as the phase-encoding dimension ($k_y$), and the head-foot direction as the slice-encoding dimension ($k_z$). No undersampling was applied along $k_x$. For each 3D echo image at each frame, 18 readouts were sampled, corresponding to a net acceleration factor of $R=1,\!147$.

To demonstrate the contribution of the proposed components, two ablation variants were evaluated. First, \method\ without joint estimation of coil sensitivity maps, in which fixed ESPIRiT maps were used instead, was evaluated and denoted as ``\methodAblationSen''. The ``\methodAblationSen'' variant was implemented by removing the sensitivity-map decoder from the Y-Net, converting the network into a U-Net with a single image-basis output. 
Second, \method\ without joint sensitivity-map estimation and without multi-encoding was evaluated by reconstructing the two echoes independently using separate training for each echo (referred to as ``\methodAblationSenEcho''). The ``\methodAblationSenEcho'' variant was implemented by generating only a frame-specific coefficient matrix $\vec{W}_{b}^{(t)} \in \Complex^{L_2 \times 1}$ from the spatial MLP and synthesizing a single-\echo\ motion-free image.

\subsubsection{In vivo animal study}
To further evaluate \method, prospectively undersampled iNAV-based dual-\echo\ 3D LGE data were acquired in five ventilated Yucatan minipigs on a 3\,T scanner (MAGNETOM Vida, Siemens Healthineers, Germany) following gadolinium administration.\cite{bratis2016iNAV-seq} 
Approval was granted by the Institutional Animal Care and Use Committee (IACUC) at QTest Labs, Columbus, OH (SPP24-010).
Data were acquired during free breathing for approximately five minutes, with 18--26 readouts per 3D echo image of each frame. Sequence parameters included echo times (TEs) of 1.39 and 2.87\,ms, repetition time (TR) of 5.0\,ms, temporal footprint of 180--260\,ms, inversion time (TI) of 220--310\,ms, flip angle of 15$^\circ$, and acceleration factors $R$ ranging from $768$ to $1,\!461$. Reconstruction matrix sizes ranged from $(128\text{--}144)\times(264\text{--}288)\times(96\text{--}112)$, with spatial resolution of $1.0 \times 1.0 \times 1.25$\,mm$^3$. The number of reconstructed frames ranged from 199 to 377.

\subsubsection{In vivo human study}
Prospectively undersampled dual-\echo\ free-breathing 3D LGE data were also acquired in eight human patients referred clinically for LGE imaging on a 3\,T scanner (MAGNETOM Vida, Siemens Healthineers, Germany).\cite{bratis2016iNAV-seq} 
Approval was granted by the Institutional Review Board (IRB) at The Ohio State University (2019H0076). Written informed consent was obtained from each patient prior to imaging.
Each dataset was acquired over approximately five minutes using an iNAV-based dual-\echo\ 3D LGE sequence with 18 readouts per 3D echo image of each frame. In five patients, the anterior-posterior direction was used as the frequency-encoding direction, with reconstruction matrix sizes of $(132\text{--}138)\times342\times(64\text{--}72)$. This readout orientation is recommended for iNAV-based sequences to suppress breathing motion artifacts but leads to inefficient acquisition. In the remaining three patients, left-right frequency encoding was used, which enabled more aggressive readout cropping along the larger left-right direction, yielding a smaller reconstruction matrix of $(130\text{--}140)\times210\times96$ and faster reconstruction. Sequence parameters were TE = 1.39 and 2.87\,ms, TR = 5.0\,ms, temporal footprint = 180\,ms, TI = 300--340\,ms, flip angle = 15--20$^\circ$, $R = 960$--$1,\!368$, and spatial resolution of 1.25--1.7\,mm, 1.25--1.7\,mm, and 1.3--1.5\,mm along the frequency-encoding, phase-encoding, and slice directions, respectively. The number of reconstructed frames ranged from 248 to 353.

In human subjects, a conventional 2D phase-sensitive inversion recovery (PSIR) short-axis LGE stack was additionally acquired as part of the clinical protocol.\cite{kellman2002PSIR} The 2D reference images had in-plane resolution of $(1.2\text{--}1.5)\times(1.2\text{--}1.5)$\,mm$^2$, slice thickness of 8\,mm, temporal footprint of 120--135\,ms, TE of 1.2\,ms, TI of 330--410\,ms, and flip angle of 40$^\circ$.

\subsection{Preprocessing and comparator reconstructions}
\subsubsection{Preprocessing}
To improve computational efficiency, the dual-\echo\ k-space data were cropped along the fully sampled readout dimension before reconstruction to exclude regions distant from the heart. 
To further reduce the computational cost of animal and human studies, joint coil compression was performed across the two echoes in two stages. First, physical-coil data were compressed to 14 virtual coils using PCA\cite{buehrer2007array} as a denoising step. Second, the 14 virtual coils were further compressed to 6 region-optimized virtual coils using the ROVir method\cite{kim2021rovir} to enhance the signal in the heart region. ESPIRiT sensitivity maps, $\vec{H}_0$, corresponding to the six virtual coils were then estimated from time-averaged k-space data of the in-phase echo.\cite{uecker2014espirit}

\subsubsection{Comparator reconstructions}
For in vivo comparison, an iNAV-based motion-corrected compressed-sensing (CS) reconstruction \cite{zeilinger2021iNAV} implemented on the scanner was used as the conventional state-of-the-art baseline for this acquisition, referred to as iNAV.
In this method, rigid-body motion estimates in the head-foot and left-right directions were derived from the 2D iNAV images and used to correct the dual-\echo\ 3D data. As a deep learning baseline, we selected LR-DIP,\cite{hamilton2023lrdip2d} a subspace-based self-supervised reconstruction method. Since LR-DIP was originally developed for 2D dynamic imaging, we extended it to 3D to obtain LR-DIP-3D, which served as an advanced deep learning comparator for the proposed 3D reconstruction task.
For offline LR-DIP-3D reconstruction, fixed ESPIRiT maps were used in all experiments. Hyperparameters, including rank = 40 and 500 training epochs, were initialized from the 2D implementation and empirically tuned for the 3D setting. In both iNAV and LR-DIP-3D, the two echoes were reconstructed independently, following their default configurations. Because iNAV produced a single motion-corrected LGE image, a representative non-outlier respiratory frame that recurred throughout the scan was selected from the motion-resolved LR-DIP-3D and \method\ reconstructions for all in vivo comparisons.


\subsubsection{Reconstruction runtime}
All \method\ reconstructions were performed on a single H100 GPU (NVIDIA, Santa Clara, CA, USA). Depending on the reconstruction matrix size, training required approximately 4--5 hours per dataset. LR-DIP-3D training required approximately 3--5 hours on the same hardware.

\subsection{Postprocessing and evaluation}
\subsubsection{Postprocessing}
For in vivo studies, post-reconstruction image intensity correction was applied in two stages. First, surface coil intensity correction (SCC) was used to compensate for the inherent spatial non-uniformity of the receive coils' sensitivity maps.\cite{lei2024SCC} Second, a reverse-ROVir map was additionally applied to compensate for intensity variation introduced by ROVir coil compression during preprocessing. These corrections allowed direct comparison between our offline reconstructions and the inline iNAV reconstruction. Fat-water separation was then performed using IDEAL-CE\cite{bydder2020PDFF} for the offline \method\ and LR-DIP-3D reconstructions. The regularization parameters in IDEAL-CE were tuned to $\mu_B = 0.01$ and $\mu_R = 0.05$ to encourage smooth phase and field maps and to minimize fat-water swap artifacts.

\subsubsection{Quantitative evaluation}
For phantom experiments, reconstructed echo images were first converted to multi-coil images using simulated coil sensitivity maps. The resulting complex-valued multi-coil reconstructed images were then compared with the noiseless reference multi-coil echo images using peak signal-to-noise ratio (PSNR$\uparrow$), defined as 
$20\log_{10} \left( \frac{\max(\vec{x})}{\|\Hvec{x} - \vec{x}\|_2 / \sqrt{N}} \right)$ (dB),
and structural similarity index measure (SSIM$\uparrow$).\cite{wang2004ssim} The metrics were computed for a randomly selected frame. 
Average PSNR and SSIM values are reported along with the standard error of the mean (SEM), where $\mathrm{SEM}$ is defined as $\mathrm{SD}/\sqrt{n}$, with $n$ denoting the number of subjects.

\subsubsection{Reader study}
For the animal and human studies, fat and water images reconstructed using iNAV, LR-DIP-3D, and \method\ were reviewed in a blinded manner by two Level-3 cardiovascular MRI-trained cardiologists. For each subject, the readers were presented with 3D LGE movies together with representative image slices on a single slide, with display order randomized across reconstruction methods. Each reader assigned an overall image-quality score ($\uparrow$) on a five-point Likert scale: 1 = non-diagnostic, 2 = poor, 3 = fair, 4 = good, and 5 = excellent. In addition to the scores, the readers were asked to select the preferred reconstruction for each subject.

For the human study, the reconstructed 3D fat and water volumes were additionally reformatted into standard short-axis views using slice-position information from the 2D PSIR LGE acquisition. 
\begin{figure*}[!t]
\centering
\includegraphics[width=0.95\textwidth]{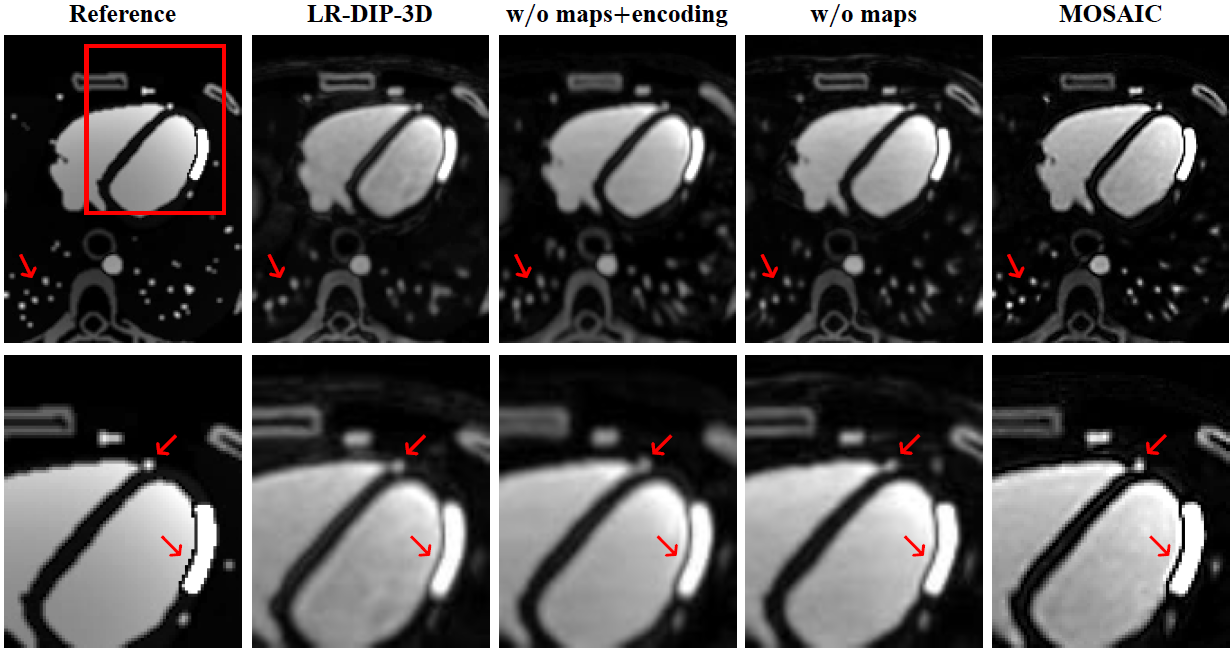}
\caption{Representative phantom results for Sub.~\#5 in \tabref{ablations_phantom}. The first column shows the reference image, the second shows LR-DIP-3D, the third and fourth show ablation variants of \method\ (``\methodAblationSenEcho'' denotes \method\ without joint sensitivity-map estimation and without multi-encoding, whereas ``\methodAblationSen'' denotes \method\ without joint sensitivity-map estimation only), and the last shows the proposed \method. The first row shows water images from a selected axial slice, and the second row shows an enlarged region indicated by the red box. Red arrows highlight regions where \method\ better preserves small structures and sharp tissue boundaries. The corresponding fat images are not shown.}
\label{fig:xcat_sub5}
\end{figure*}

\begin{table}[!t]
\footnotesize
\centering
\setlength{\tabcolsep}{2.3pt}
\renewcommand{\arraystretch}{1.0}
\begin{tabular}{l|cccc}
\toprule
Sub. & LR-DIP-3D & \methodAblationSenEcho & \methodAblationSen & \method \\
\midrule
1 & 39.17$/$0.925 & 43.57$/$0.969 & 44.14$/$0.980 & \textbf{46.54}$/$\textbf{0.986} \\
2 & 42.82$/$0.965 & 43.34$/$0.975 & 43.86$/$0.977 & \textbf{46.38}$/$\textbf{0.985} \\
3 & 40.27$/$0.965 & 41.89$/$0.969 & 43.90$/$0.975 & \textbf{48.32}$/$\textbf{0.988} \\
4 & 40.10$/$0.962 & 41.72$/$0.971 & 43.93$/$0.974 & \textbf{45.54}$/$\textbf{0.977} \\
5 & 41.96$/$0.968 & 43.01$/$0.975 & 43.24$/$0.977 & \textbf{48.20}$/$\textbf{0.989} \\
6 & 40.41$/$0.976 & 42.51$/$0.967 & 43.53$/$0.979 & \textbf{48.99}$/$\textbf{0.990} \\
\midrule
Avg. & 40.79$/$0.960 & 42.67$/$0.971 & 43.77$/$0.977 & \textbf{47.33}$/$\textbf{0.986} \\
SEM & 0.548$/$0.0073 & 0.311$/$0.0014 & 0.132$/$0.0009 & 0.554$/$0.0019 \\
\bottomrule
\end{tabular}
\caption{PSNR$/$SSIM values from the phantom study. ``\methodAblationSenEcho'' denotes \method\ without joint sensitivity-map estimation and without multi-encoding, whereas ``\methodAblationSen'' denotes \method\ without joint sensitivity-map estimation only.}
\label{tab:ablations_phantom}
\end{table}

\section{Results}
\subsection{3D LGE phantom study}
We first evaluated \method\ on six MRXCAT phantoms and compared its performance with LR-DIP-3D. Ablation studies were also performed to assess the individual contributions of the proposed multi-encoding and integrated sensitivity-map estimation. Quantitative results in terms of PSNR and SSIM are summarized in \tabref{ablations_phantom}. 
Across the six anatomically distinct phantoms, LR-DIP-3D yielded the lowest reconstruction quality, with a mean PSNR$/$SSIM of 40.79~dB$/$0.960. Among the ablation variants, removing both joint sensitivity-map estimation and multi-encoding (\method\ \methodAblationSenEcho) yielded an average PSNR$/$SSIM of 42.67~dB$/$0.971, while adding joint echo processing but still fixing the sensitivity maps (\method\ \methodAblationSen) improved the average PSNR$/$SSIM to 43.77~dB$/$0.977. Full \method\ achieved the best performance, with the highest average PSNR and SSIM of 47.33~dB and 0.986, respectively.

Representative phantom results for Sub.~\#5 in \tabref{ablations_phantom} are shown in \figref{xcat_sub5}. Compared with LR-DIP-3D, the ablation variants showed improved image sharpness and anatomical delineation. The full \method\ provided the best visual quality, with better preservation of small structures and sharper tissue boundaries in both the full field-of-view images in the first row and the corresponding zoomed views in the second row. These differences are further highlighted in the enlarged views, where the structures indicated by the arrows are more clearly resolved using \method. Supplementary Video S1 provides slice-by-slice fat and water movies corresponding to \figref{xcat_sub5}. 

\begin{figure*}[!t]
\centering
\includegraphics[width=0.75\textwidth]{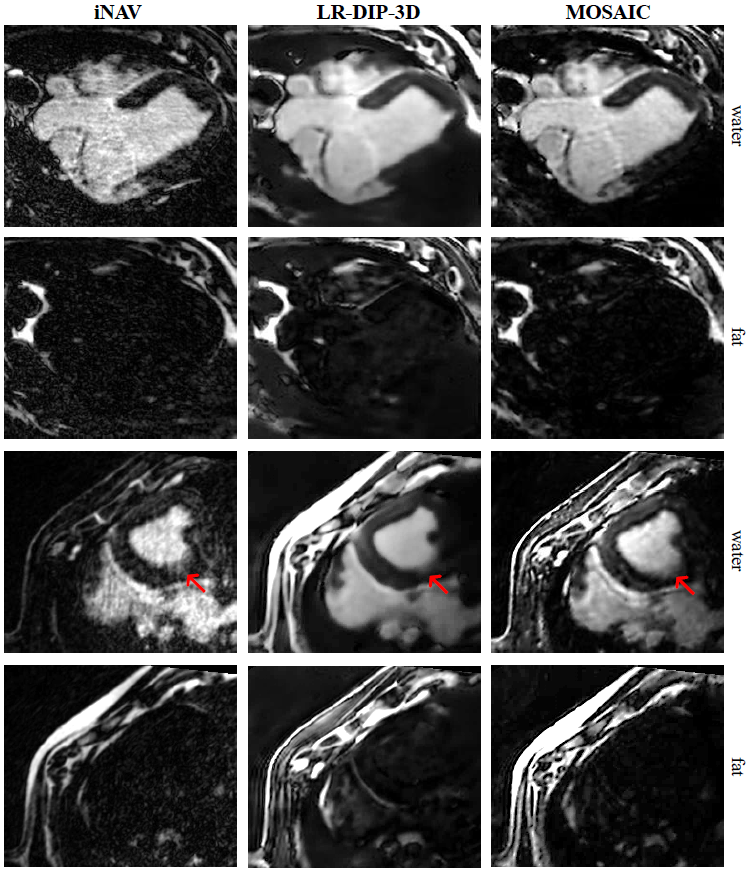}
\caption{Representative animal results for Sub.~\#1 in \tabref{scores_pigs}. The first row shows a selected axial slice, and the third row shows a reformatted short-axis slice. For each method (iNAV, LR-DIP-3D, and \method), water and fat images are displayed in adjacent rows. The red arrows in the third row highlight artifacts (iNAV) and blurring (LR-DIP-3D) in the competing methods.}
\label{fig:pig_sub1}
\end{figure*}

\begin{table}[!t]
\footnotesize
\centering
\setlength{\tabcolsep}{2.5pt}
\renewcommand{\arraystretch}{1.0}
\begin{tabular}{l|cccc}
\toprule
Sub. & iNAV & LR-DIP-3D & \method & Preferred \\
\midrule
1 & 3.00 & 2.50 & \textbf{3.75} & \method \\
2 & 2.75 & 3.00 & \textbf{4.00} & \method \\
3 & 4.00 & 2.50 & \textbf{4.25} & \method, iNAV \\
4 & \textbf{4.25} & 3.25 & 3.50 & iNAV \\
5 & 3.50 & 2.75 & \textbf{3.75} & \method \\
\midrule
Avg. & 3.50 & 2.80 & \textbf{3.85} & -- \\
SEM & 0.29 & 0.15 & 0.13 & -- \\
Count & 3 & 0 & \textbf{7} & -- \\
\bottomrule
\end{tabular}
\caption{Blinded image-quality scores and preferred-reconstruction votes for the animal study. Subject-wise scores are averaged across two readers.}
\label{tab:scores_pigs}
\end{table}

\subsection{In vivo animal study}
We next evaluated \method\ using prospectively acquired animal data from five pigs and compared its reconstruction performance with inline iNAV and LR-DIP-3D. Representative qualitative results are shown in \figref{pig_sub1}, and blinded reader-study scores are summarized in \tabref{scores_pigs}. 
\figref{pig_sub1} shows the fat and water images for Sub.~\#1 in \tabref{scores_pigs}. 
In the axial view, \method\ provides a more favorable balance between artifact and noise suppression and preservation of fine details.
In the short-axis view, as highlighted by the red arrows, iNAV shows residual motion artifacts while LR-DIP-3D exhibits excessive blurring.
Supplementary Video S2 shows the corresponding slice-by-slice fat and water movies.

These visual findings were consistent with the blinded reader study performed by two cardiologists, as summarized in \tabref{scores_pigs}. For each subject, the table reports the average image-quality score across the two readers and the preferred-reconstruction vote. When the two readers selected different methods, both preferences are listed. The final rows report the mean image-quality score with SEM, as well as the total number of preferred-reconstruction votes for each method across all five subjects. Overall, \method\ achieved the highest mean image-quality score ($3.85$) and received the largest number of preferred-reconstruction votes (7$/$10).


\begin{figure*}[!t]
\centering
\includegraphics[width=0.95\textwidth]{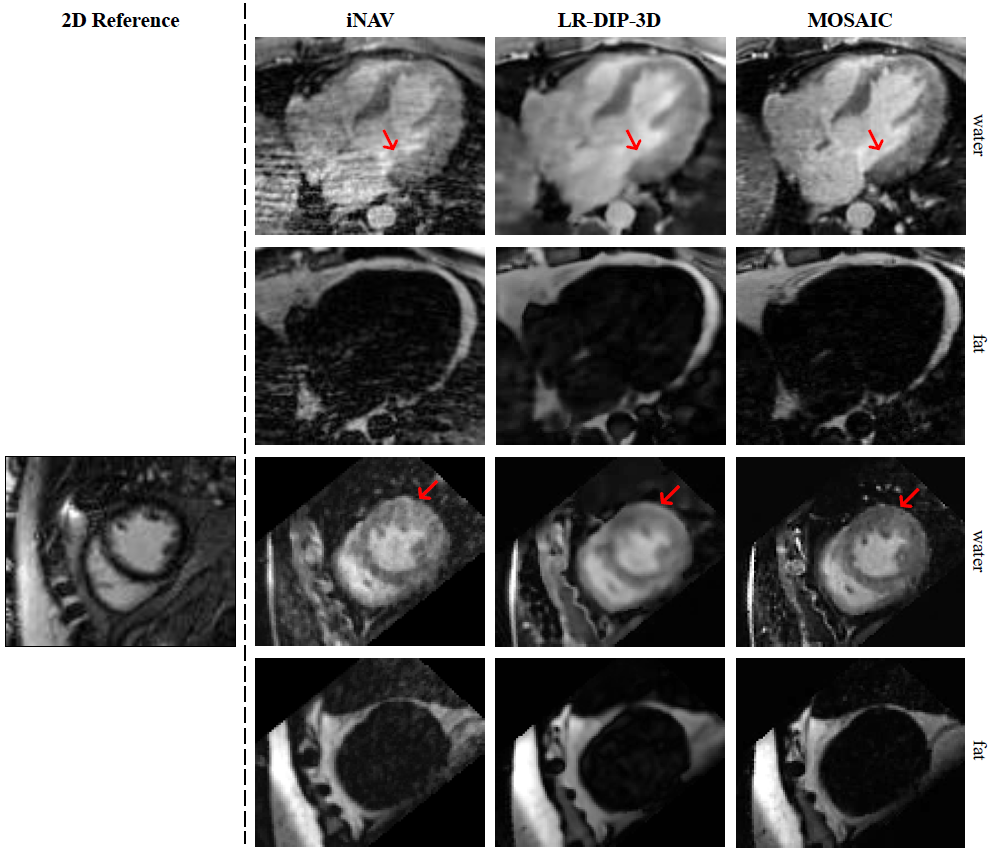}
\caption{Representative human results for Sub.~\#2 in \tabref{scores_humans}, with left-right readout direction. The first row shows a selected axial slice, and the third row shows a slice matched to the 2D short-axis reference (left-most image). For each 3D method (iNAV, LR-DIP-3D, and \method), water and fat images are displayed in adjacent rows. The red arrows in the first row highlight a region where iNAV shows pronounced residual artifacts, and the red arrows in the third row highlight a region where \method\ better preserves both the boundary sharpness and expected uniformity of the myocardium.}
\label{fig:pt_sub2}
\end{figure*}

\begin{figure*}[!t]
\centering
\includegraphics[width=0.95\textwidth]{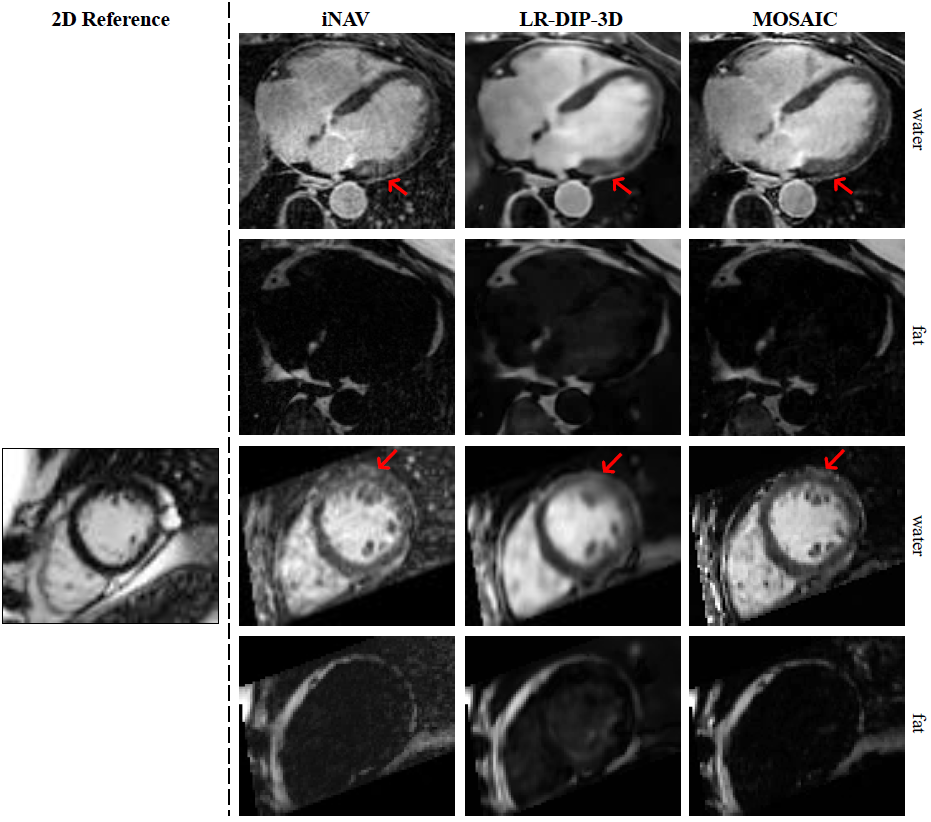}
\caption{Representative human results for Sub.~\#3 in \tabref{scores_humans}, with anterior-posterior readout direction. The first row shows a selected axial slice, and the third row shows a slice matched to the 2D short-axis reference (left-most image). For each 3D method (iNAV, LR-DIP-3D, and \method), water and fat images are displayed in adjacent rows. The red arrows highlight regions where \method\ better preserves both the boundary sharpness and expected uniformity of the myocardium.}
\label{fig:pt_sub3}
\end{figure*}

\subsection{In vivo human study}
\begin{figure*}[!t]
\centering
\includegraphics[width=0.95\textwidth]{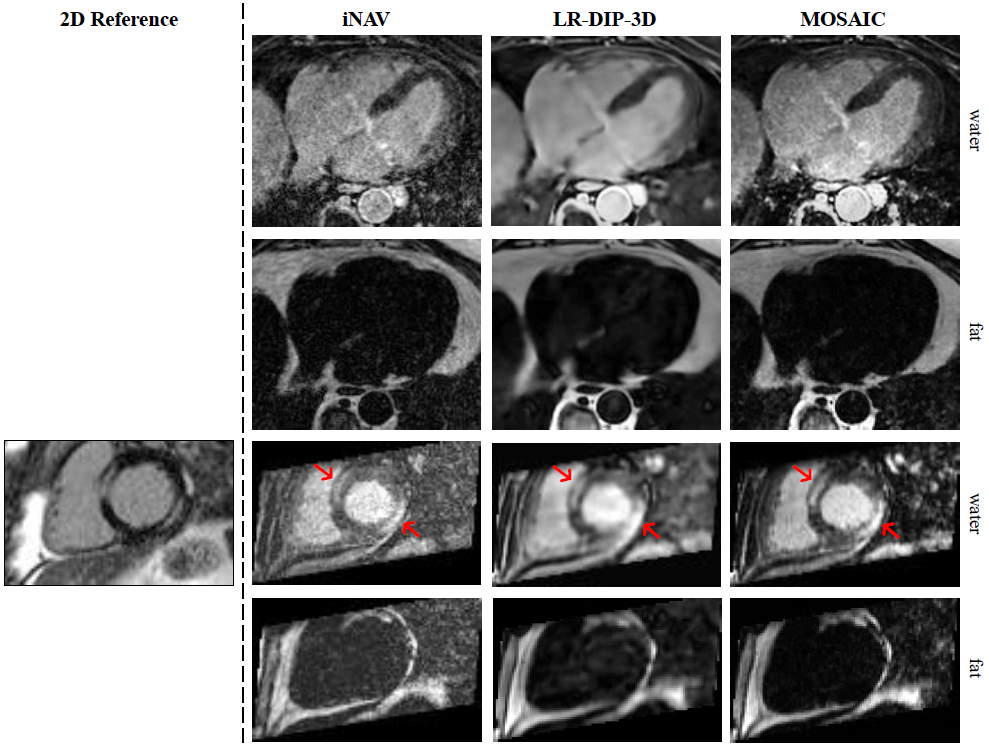}
\caption{Representative human results for Sub.~\#6 in \tabref{scores_humans}, acquired with anterior-posterior readout direction and showing visible myocardial scar. The first row shows a selected axial slice, and the third row shows a slice matched to the 2D short-axis reference (left-most image), in which the scar is visible. For each 3D method (iNAV, LR-DIP-3D, and \method), water and fat images are displayed in adjacent rows. The red arrows highlight regions in the short-axis view where \method\ provides clearer scar depiction than iNAV and LR-DIP-3D.}
\label{fig:pt_sub6_scar}
\end{figure*}

Finally, we evaluated \method\ on eight prospective clinical patient datasets against inline iNAV and offline LR-DIP-3D. 
Representative qualitative results are shown in \figsref{pt_sub2}, \fignoref{pt_sub3}, and \fignoref{pt_sub6_scar}, and blinded reader-study scores are summarized in \tabref{scores_humans}.
\figref{pt_sub2} and \figref{pt_sub3} show representative results for Sub.~\#2 and \#3 in \tabref{scores_humans}, respectively. The same qualitative trends observed in the animal study were also seen in the patient data. In both axial and short-axis views, \method\ provided improved myocardial structural conspicuity and clearer tissue boundaries than the comparison methods, as highlighted by the arrows. In \figref{pt_sub2}, which was acquired using a left-right readout direction, both iNAV and LR-DIP-3D showed marked degradation in image quality, whereas \method\ maintained sufficient structural depiction. 
An additional case with visible scar, corresponding to Sub.~\#6 in \tabref{scores_humans}, is shown in \figref{pt_sub6_scar}. 
Compared with iNAV and LR-DIP-3D, \method\ provided clearer scar depiction in the short-axis view, as highlighted by arrows.

The blinded reader study in \tabref{scores_humans} further supports these observations. Across the patient cohort, \method\ achieved the highest mean image-quality score (4.06/5) and was selected as the preferred reconstruction in 13 of 16 reader assessments, outperforming both iNAV and LR-DIP-3D.


Slice-by-slice results corresponding to \figsref{pt_sub2} to \fignoref{pt_sub6_scar} are provided in Supplementary Videos S3--S5, respectively. 
These videos further demonstrate the spatial extent of reconstruction artifacts, particularly for the subject shown in \figref{pt_sub3}, where iNAV artifacts are more pronounced across multiple slices.

In addition to the comparisons with iNAV and LR-DIP-3D, Supplementary Figure S1 presents ablation results for the same subject shown in \figref{pt_sub3}, and Supplementary Video S6 presents the corresponding slice-by-slice ablation results.
Supplementary Video S7 further presents frame-by-frame results from \method's\ motion-resolved reconstruction across eight contiguous frames for the same subject shown in \figref{pt_sub2}.

\begin{table}[!t]
\footnotesize
\centering
\setlength{\tabcolsep}{2.5pt}
\renewcommand{\arraystretch}{1.0}
\begin{tabular}{l|cccc}
\toprule
Sub. & iNAV & LR-DIP-3D & \method & Preferred \\
\midrule
1 & \textbf{4.75} & 3.25 & 4.00 & iNAV \\
2 & 2.25 & 2.50 & \textbf{3.25} & \method \\
3 & 4.00 & 3.00 & \textbf{4.75} & \method \\
4 & 3.75 & 2.75 & \textbf{4.25} & \method \\
5 & 2.00 & 2.00 & \textbf{3.25} & \method \\
6 & 4.00 & 2.75 & \textbf{4.75} & \method \\
7 & 3.75 & 3.00 & \textbf{4.25} & \method \\
8 & \textbf{4.25} & 3.75 & 4.00 & \method, iNAV\\
\midrule
Avg. & 3.59 & 2.88 & \textbf{4.06} & -- \\
SEM & 0.34 & 0.18 & 0.20 & -- \\
Count & 3 & 0 & \textbf{13} & -- \\
\bottomrule
\end{tabular}
\caption{Blinded image-quality scores and preferred-reconstruction votes for the human study. Subject-wise scores are averaged across two readers.}
\label{tab:scores_humans}
\end{table}



\section{Discussion} 
The phantom and in vivo results indicate that \method\ improves reconstruction quality relative to LR-DIP-3D and inline iNAV reconstructions. In the phantom experiments, \method\ achieved the highest PSNR and SSIM in all cases (\tabref{ablations_phantom}). Consistent with these quantitative improvements, the representative example in \figref{xcat_sub5} shows better preservation of small structures and more clearly defined tissue boundaries with \method\ than with LR-DIP-3D, as indicated by the arrows. These findings indicate that the explicit motion modeling and multi-encoding in \method\ provide stronger constraints and richer modeling compared to subspace-based methods, such as LR-DIP-3D. 


The ablation experiments demonstrate the contribution of the main components of \method. Removing both joint echo processing (multi-encoding) and integrated sensitivity-map estimation (\method\ \methodAblationSenEcho) reduced performance substantially, while adding joint echo processing but still fixing the sensitivity maps (\method\ \methodAblationSen) provided only partial recovery, as evident from \tabref{ablations_phantom}, \figref{xcat_sub5}, and Supplementary Figure S1. Together, these findings indicate that both components contribute meaningfully to the final reconstruction quality. The gain from joint echo processing is consistent with the use of a shared spatial basis across echoes, which allows the model to exploit inter-\echo\ redundancy and additionally regularize the highly ill-posed multi-\echo\ 3D LGE reconstruction problem. The additional gain from integrated sensitivity-map estimation supports that jointly estimating smooth coil sensitivity maps within the reconstruction is beneficial when calibration information from time-averaged k-space data is limited or noisy.
Finally, the relatively poor performance of ``\method\ \methodAblationSenEcho'' further supports that without the key innovations introduced by \method, the cine-specific ML-DIP or similar methods are not competitive for multi-\echo\ 3D LGE.

The in vivo studies also revealed distinct failure modes for the comparison methods. Inline iNAV reconstruction often produced relatively sharp images, but with elevated noise and residual motion artifacts, which can hinder clinical interpretation. In contrast, LR-DIP-3D suppressed noise more strongly but tended to over-smooth fine anatomical structure, potentially reducing the conspicuity of clinically relevant scar patterns. Across both animal and human datasets, \method\ provided a more favorable tradeoff between detail preservation and artifact suppression, which is particularly important for LGE-based scar assessment. The advantage of \method\ is reflected in the visual comparisons in \figsref{pig_sub1}, \fignoref{pt_sub3}, and \fignoref{pt_sub6_scar}, as well as in the reader-study results in \tabsref{scores_pigs} and \tabnoref{scores_humans}. 
Owing to the small number of animal datasets, formal statistical testing was not performed for the animal reader-study scores, and these results were therefore interpreted descriptively. 
The patient case shown in \figref{pt_sub2} is particularly noteworthy: under a left-right readout direction, both iNAV and LR-DIP-3D degraded markedly, whereas \method\ remained substantially more robust because of its explicit motion modeling. 

Although \method\ achieved the highest mean scores and the largest number of preferred-reconstruction votes in the reader-study results, iNAV was still preferred in some cases. This indicates that, under relatively favorable motion conditions, such as regular breathing and normal heartbeats, the 2D rigid-body motion correction provided by iNAV may be sufficient, resulting in overall modest differences between methods. 
In contrast to iNAV, \method\ reconstructs a 3D multi-\echo\ LGE frame from each heartbeat, as shown in Supplementary Video S7.  
This motion-resolved series provides multiple candidate frames for downstream clinical assessment.
Although not the focus of this work, this redundancy afforded by \method\ can enable selecting a more ``typical'' frame and avoiding outlier frames contaminated by bulk motion or arrhythmias.  

For the human study, reformatted short-axis images from the clinical two-dimensional PSIR acquisition are also presented as the left-most images in \figsref{pt_sub2}, \fignoref{pt_sub3}, and \fignoref{pt_sub6_scar}. These images were provided to the readers for all datasets to assist in interpretation as part of the clinical assessment process. Because the 2D and 3D acquisitions were performed at different times after gadolinium administration, the image contrast was not expected to match. Furthermore, differences in respiratory and cardiac phase between the two acquisitions precluded exact slice-to-slice correspondence, as reflected in \figsref{pt_sub2}, \fignoref{pt_sub3}, and \fignoref{pt_sub6_scar}.

A separate limitation arises from the post-reconstruction step of fat-water separation. In the current pipelines for \method\ and LR-DIP-3D, dual-\echo\ images are first reconstructed, after which fat and water images are obtained using the external IDEAL-CE method. We observed occasional visible fat-water swaps produced by IDEAL-CE, for example, in the chest-wall fat layer for LR-DIP-3D in \figref{pig_sub1}. This suggests that some image-quality errors may originate not only from the reconstruction itself but also from the downstream fat-water separation step. A natural direction for future work is therefore to incorporate a Dixon signal model \cite{ma2008dixon} directly into \method\ so that echo reconstruction and fat-water separation are performed jointly in an end-to-end framework.

This study has several other limitations. First, the sample size is modest, and larger studies are needed to further assess generalizability. Second, further validation is needed in larger and more diverse LGE cohorts, including patients with arrhythmias, to establish the robustness of the proposed method. 
Third, although \method\ was compared with the established inline iNAV reconstruction and a 3D version of LR-DIP as a self-supervised deep learning baseline, additional reconstruction methods may be considered in future studies as more approaches become available for highly undersampled free-breathing multi-\echo\ 3D LGE. 
Fourth, the reconstruction time remains relatively long, which may limit immediate practical deployment. Computational acceleration strategies, including network pretraining using coarse initial reconstructions and progressive coarse-to-fine optimization, may help reduce reconstruction time. Future work will therefore focus on larger clinical validation, computational acceleration, and extension to related free-breathing volumetric applications, such as free-breathing chest MR angiography.

\section{Conclusion}
We presented \method, a motion-robust reconstruction framework for highly undersampled free-breathing 3D LGE MRI. The proposed method combines frame-specific motion modeling with a shared-basis representation of dual-\echo\ image content and integrated sensitivity-map estimation within a single self-supervised reconstruction framework. By avoiding respiratory binning, \method\ explicitly models underlying motion across heartbeats, while the shared image basis further constrains the dual-\echo\ reconstruction and the joint map-estimation strategy reduces errors from limited calibration data. Phantom and in vivo results support that these design choices improve artifact suppression and detail preservation relative to the comparison methods. 
\section*{Acknowledgments}
The authors acknowledge the Ohio Supercomputer Center for providing the computational resources used in this work. The authors thank Juliet Varghese for preliminary image-quality feedback, Syed Murtaza Arshad for assistance with MRXCAT simulations, Mark Vornehm for sharing the M-DIP code, and Jesse Hamilton for sharing the LR-DIP-2D code.

\subsection*{Author contributions}
\noindent
\textbf{Muhammad A. Sultan:}
Conceptualization, Methodology, Software, Data curation,
Investigation, Validation, Formal analysis, Visualization,
Writing--original draft, Project administration;
\textbf{Yingmin Liu:}
Data curation, Software, Writing--review \& editing;
\textbf{Katherine Binzel:}
Data curation, Writing--review \& editing;
\textbf{Katarzyna E. Gil:}
Validation, Writing--review \& editing;
\textbf{Karolina M. Zareba:}
Validation, Writing--review \& editing;
\textbf{Rizwan Ahmad:}
Conceptualization, Methodology, Investigation, Supervision,
Funding acquisition, Resources, Writing--review \& editing.


\subsection*{Data availability statement}
The code and representative data are available at \url{https://github.com/OSU-MR/MOSAIC}.


\bibliography{main_MRM}

\begin{thebibliography}{10}

\bibitem{kim2000lge}
Kim RJ, Wu~E, Rafael A, et al. The use of contrast-enhanced magnetic resonance imaging to identify reversible myocardial dysfunction.  {\it N Engl J Med. }2000;343(20):1445--1453.

\bibitem{piehler2013FB-LGE}
Piehler KM, Wong TC, Puntil KS, et al. Free-breathing, motion-corrected late gadolinium enhancement is robust and extends risk stratification to vulnerable patients.  {\it Circ Cardiovasc Imaging. }2013;6(3):423--432.

\bibitem{toupin2022lge}
Toupin S, Pezel T, Bustin A, Cochet H. Whole-heart high-resolution late gadolinium enhancement: techniques and clinical applications.  {\it J Magn Reson Imaging. }2022;55(4):967--987.

\bibitem{kellman2002PSIR}
Kellman P, Arai AE, McVeigh ER, Aletras AH. Phase-sensitive inversion recovery for detecting myocardial infarction using gadolinium-delayed hyperenhancement.  {\it Magn Reson Med. }2002;47(2):372--383.

\bibitem{abdula2014psir}
Abdula G, S{\"o}rensson P, Lundin M, et al. Synthetic phase-sensitive inversion-recovery late gadolinium enhancement from postcontrast {T1} mapping shows excellent agreement with conventional {PSIR-LGE} for diagnosing myocardial scar.  {\it J Cardiovasc Magn Reson. }2014;16(Suppl 1):P213.

\bibitem{peters2009lge3d}
Peters DC, Appelbaum EA, Nezafat R, et al. Left ventricular infarct size, peri-infarct zone, and papillary scar measurements: a comparison of high-resolution {3D} and conventional {2D} late gadolinium enhancement cardiac {MR}.  {\it J Magn Reson Imaging. }2009;30(4):794--800.

\bibitem{kino2009threeD}
Kino A, Zuehlsdorff S, Sheehan JJ, et al. Three-dimensional phase-sensitive inversion-recovery turbo {FLASH} sequence for the evaluation of left ventricular myocardial scar.  {\it AJR Am J Roentgenol. }2009;193(5):W381--W388.

\bibitem{morita2013comparison}
Morita K, Utsunomiya D, Oda S, et al. Comparison of {3D} phase-sensitive inversion-recovery and {2D} inversion-recovery {MRI} at {3.0 T} for the assessment of late gadolinium enhancement in patients with hypertrophic cardiomyopathy.  {\it Acad Radiol. }2013;20(6):752--757.

\bibitem{morsbach2016comp}
Morsbach F, Gordic S, Gruner C, et al. Quantitative comparison of {2D} and {3D} late gadolinium enhancement {MR} imaging in patients with {Fabry} disease and hypertrophic cardiomyopathy.  {\it Int J Cardiol. }2016;217:167--173.

\bibitem{pennig2020lge}
Pennig L, Lennartz S, Wagner A, et al. Clinical application of free-breathing {3D} whole-heart late gadolinium enhancement cardiovascular magnetic resonance with high isotropic spatial resolution using {Compressed SENSE}.  {\it J Cardiovasc Magn Reson. }2020;22(1):89.

\bibitem{saranathan2004dNAV}
Saranathan M, Rochitte CE, Foo TKF. Fast, three-dimensional free-breathing {MR} imaging of myocardial infarction: a feasibility study.  {\it Magn Reson Med. }2004;51(5):1055--1060.

\bibitem{nguyen2008dNAV3d}
Nguyen TD, Spincemaille P, Weinsaft JW, et al. A fast navigator-gated {3D} sequence for delayed-enhancement {MRI} of the myocardium: comparison with breath-hold {2D} imaging.  {\it J Magn Reson Imaging. }2008;27(4):802--808.

\bibitem{prieto2015dNAVretro}
Prieto C, Doneva M, Usman M, et al. Highly efficient respiratory motion-compensated free-breathing coronary {MRA} using golden-step {Cartesian} acquisition.  {\it J Magn Reson Imaging. }2015;41(3):738--746.

\bibitem{munoz2020binning}
Munoz C, Bustin A, Neji R, et al. Motion-corrected {3D} whole-heart water-fat high-resolution late gadolinium enhancement cardiovascular magnetic resonance imaging.  {\it J Cardiovasc Magn Reson. }2020;22(1):53.

\bibitem{zeilinger2022binning}
Zeilinger MG, Kunze KP, Munoz C, et al. Non-rigid motion-corrected free-breathing {3D} myocardial {Dixon LGE} imaging in a clinical setting.  {\it Eur Radiol. }2022;32(7):4340--4351.

\bibitem{bratis2016iNAV-seq}
Bratis K, Henningsson M, Grigoratos C, et al. Image-navigated three-dimensional late gadolinium enhancement cardiovascular magnetic resonance imaging: feasibility and initial clinical results.  {\it J Cardiovasc Magn Reson. }2017;19(1):97.

\bibitem{zeilinger2021iNAV}
Zeilinger MG, Wiesm{\"u}ller M, Forman C, et al. {3D Dixon} water-fat {LGE} imaging with image navigator and compressed sensing in cardiac {MRI}.  {\it Eur Radiol. }2021;31(6):3951--3961.

\bibitem{holtackers2022lge}
Holtackers RJ, Emrich T, Botnar RM, Kooi ME, Wildberger JE, Kreitner KF. Late gadolinium enhancement cardiac magnetic resonance imaging: from basic concepts to emerging methods.  {\it Rofo. }2022;194(5):491--504.

\bibitem{hopman2024dNAVvsiNAV}
Hopman LHGA, Sol{\'i}s-Lemus JA, Hofman MBM, et al. Performance of image-navigated and diaphragm-navigated {3D} late gadolinium-enhanced cardiac {MRI} for the assessment of atrial fibrosis.  {\it Radiol Cardiothorac Imaging. }2024;6(2):e230172.

\bibitem{ma2008dixon}
Ma~J. Dixon techniques for water and fat imaging.  {\it J Magn Reson Imaging. }2008;28(3):543--558.

\bibitem{hamilton2023lrdip2d}
Hamilton JI, Truesdell W, Galizia M, Burris N, Agarwal P, Seiberlich N. A low-rank deep image prior reconstruction for free-breathing ungated spiral functional {CMR} at {0.55 T} and {1.5 T}.  {\it MAGMA. }2023;36(3):451--464.

\bibitem{hamilton2024lrdip3d}
Hamilton J, Da~Cruz GL, Seiberlich N. {3D} free-breathing ungated cine imaging at {1.5 T} and {0.55 T} using a time- and partition-dependent deep image prior.  {\it J Cardiovasc Magn Reson. }2024;26:100122.

\bibitem{sultan2025discus}
Sultan MA, Chen C, Liu Y, Gil K, Zareba K, Ahmad R. An unsupervised method for {MRI} recovery: deep image prior with structured sparsity.  {\it MAGMA. }2025;38(5):859--871.

\bibitem{zou2022moco-storm}
Zou Q, Torres LA, Fain SB, Higano NS, Bates AJ, Jacob M. Dynamic imaging using motion-compensated smoothness regularization on manifolds ({MoCo-SToRM}).  {\it Phys Med Biol. }2022;67(14):144001.

\bibitem{vornehm2025mdip}
Vornehm M, Chen C, Sultan MA, et al. Multi-dynamic deep image prior for cardiac {MRI}.  {\it Magn Reson Med. }2025;94(6):2668--2679.

\bibitem{chen2025mldip}
Chen C, Vornehm M, Bu~Z, et al. A multi-dynamic low-rank deep image prior ({ML-DIP}) for {3D} real-time cardiovascular {MRI}.  {\it J Cardiovasc Magn Reson. }2026;28:102015.

\bibitem{uecker2014espirit}
Uecker M, Lai P, Murphy MJ, et al. {ESPIRiT}: an eigenvalue approach to autocalibrating parallel {MRI}: where {SENSE} meets {GRAPPA}.  {\it Magn Reson Med. }2014;71(3):990--1001.

\bibitem{hamilton2025multifrequency_timeDIP}
Hamilton JI, Cruz G, Truesdell W, Agarwal P, Rashid I, Seiberlich N. Multifrequency time-dependent deep image prior for real-time free-breathing cardiac imaging.  {\it NMR Biomed. }2025;38(9):e70114.

\bibitem{leynes2024nlinv}
Leynes AP, Deveshwar N, Nagarajan SS, Larson PEZ. Scan-specific self-supervised {Bayesian} deep nonlinear inversion for undersampled {MRI} reconstruction.  {\it IEEE Trans Med Imaging. }2024;43(6):2358--2369.

\bibitem{sultan2026mledips}
Sultan MA, Liu Y, Binzel K, et al. A motion-robust dual-echo {3D LGE} reconstruction framework.  {\it J Cardiovasc Magn Reson. }2026;28(Suppl 1):102628.
\newblock Conference abstract.

\bibitem{jaderberg2015warp}
Jaderberg M, Simonyan K, Zisserman A, Kavukcuoglu K. Spatial transformer networks.  {\it Advances in Neural Information Processing Systems. }2015;28:2017--2025.

\bibitem{mrxcat_wissmann2014}
Wissmann L, Santelli C, Segars WP, Kozerke S. {MRXCAT}: realistic numerical phantoms for cardiovascular magnetic resonance.  {\it J Cardiovasc Magn Reson. }2014;16:63.

\bibitem{joshi2022sampling}
Joshi M, Pruitt A, Chen C, Liu Y, Ahmad R. Pseudo-random {Cartesian} sampling for dynamic {MRI}: technical report v1.0.  {\it arXiv preprint arXiv:2206.03630. }2022;.

\bibitem{buehrer2007array}
Buehrer M, Pruessmann KP, Boesiger P, Kozerke S. Array compression for {MRI} with large coil arrays.  {\it Magn Reson Med. }2007;57(6):1131--1139.

\bibitem{kim2021rovir}
Kim D, Cauley SF, Nayak KS, Leahy RM, Haldar JP. Region-optimized virtual ({ROVir}) coils: localization and/or suppression of spatial regions using sensor-domain beamforming.  {\it Magn Reson Med. }2021;86(1):197--212.

\bibitem{lei2024SCC}
Lei X, Schniter P, Chen C, Sultan MA, Ahmad R. Surface coil intensity correction for {MRI}.  {\it Proceedings of the IEEE International Symposium on Biomedical Imaging. }2024;:1--5.

\bibitem{bydder2020PDFF}
Bydder M, Ghodrati V, Gao Y, Robson MD, Yang Y, Hu~P. Constraints in estimating the proton density fat fraction.  {\it Magn Reson Imaging. }2020;66:1--8.

\bibitem{wang2004ssim}
Wang Z, Bovik AC, Sheikh HR, Simoncelli EP. Image quality assessment: from error visibility to structural similarity.  {\it IEEE Trans Image Process. }2004;13(4):600--612.

\end{thebibliography}

\clearpage 

\section*{Supporting Information}\label{sec:supp}
The following supporting information is available as part of the online article:

\noindent 
\textbf{Supplementary Figure S1}: Representative human ablation results corresponding to Figure 5 in the main manuscript. The first row shows a selected axial slice, and the third row shows a slice matched to the 2D short-axis reference (left-most image). For each 3D method (MOSAIC w/o maps+encoding, MOSAIC w/o maps, and the full proposed MOSAIC), water and fat images are displayed in adjacent rows.

\noindent 
\textbf{Supplementary Video S1}: Representative whole-heart phantom results corresponding to Figure 2 in the main manuscript. The first column shows reference, the second shows LR-DIP-3D, the third and fourth show ablations of MOSAIC (MOSAIC w/o maps+encoding and MOSAIC w/o maps), and the last shows the full proposed MOSAIC. Slice-by-slice water and fat images are displayed in adjacent rows. 

\noindent 
\textbf{Supplementary Video S2}: Representative whole-heart animal results corresponding to Figure 3 in the main manuscript. For each method (iNAV, LR-DIP-3D, and MOSAIC), slice-by-slice water and fat images are displayed in adjacent rows. 

\noindent 
\textbf{Supplementary Video S3}: Representative whole-heart human results corresponding to Figure 4 in the main manuscript. For each method (iNAV, LR-DIP-3D, and MOSAIC), slice-by-slice water and fat images are displayed in adjacent rows. 

\noindent 
\textbf{Supplementary Video S4}: Representative whole-heart human results corresponding to Figure 5 in the main manuscript. For each method (iNAV, LR-DIP-3D, and MOSAIC), slice-by-slice water and fat images are displayed in adjacent rows. 

\noindent 
\textbf{Supplementary Video S5}: Representative whole-heart human results corresponding to Figure 6 in the main manuscript. For each method (iNAV, LR-DIP-3D, and MOSAIC), slice-by-slice water and fat images are displayed in adjacent rows. 

\noindent 
\textbf{Supplementary Video S6}: Representative whole-heart human ablation results corresponding to Supplementary Figure S1. For each method (w/o maps+encoding, w/o maps, and the full proposed MOSAIC), slice-by-slice water and fat images are displayed in adjacent rows. 

\noindent 
\textbf{Supplementary Video S7}: Representative frame-by-frame human results corresponding to Figure 4 in the main manuscript. From MOSAIC’s motion-resolved reconstruction, eight contiguous frames are shown. Water and fat images are displayed in adjacent columns.

\clearpage

\end{document}